\documentclass[twocolumn,showpacs,preprintnumbers,floatfix,aps]{revtex4}

\usepackage{graphicx}
\usepackage{dcolumn}
\usepackage{bm}

\begin{document}

\preprint{IRB-ZEF/52-2003}

\title{CERN Axion Solar Telescope as a probe of large extra dimensions}

\author{R. Horvat} 
\author{M. Kr\v{c}mar} 
\author{B. Laki\'{c}} 
\affiliation{Rudjer Bo\v{s}kovi\'{c} Institute, P.O.Box 180, 10002 Zagreb,
             Croatia}

\date{\today}

\begin{abstract}
 We explore the potential of the CERN Axion Solar Telescope (CAST)
 for testing the presence of large extra dimensions. The CAST experiment has
 originally been proposed to search for solar axions with a sensitivity
 supposed to provide a limit on the axion-photon coupling
 $g_{a\gamma\gamma} \alt 5 \times 10^{-11}\;{\rm GeV^{-1}}$ or even lower.
 The expected bound on the coupling constant is by a 
 factor of ten more stringent than the
 current experimental results. This bound extends for the first time beyond 
 the limit dictated by astrophysical considerations. As a tuning
 experiment planning to explore the axion mass region up
 to about 1 eV, CAST would also be sensitive to the existence of 
 Kaluza-Klein massive states. Therefore, the detection of X-rays at least
 at two pressures may be the signature of large extra dimensions. From this
 requirement we find that CAST may test (two) large extra dimensions with a
 (common) compactification radius $R$ down to around 250 nm if
 $m_{\rm PQ} < 1/(2R)$,
 and down to around 370 nm if $1/(2R) < m_{\rm PQ}$, where $m_{\rm PQ}$ is
 the Peccei-Quinn mass.

\end{abstract}


\maketitle

\section{\label{1}INTRODUCTION}

A few years ago an approach was put forward \cite{Ark98}
in which heavy mass scales in four dimensions could be replaced by lighter
mass scales in higher dimensions. Such a class of theories is nowadays 
conventionally
considered in the context of the brane paradigm. In one class of models
extra dimensions are felt only by gravity (as well as  other fields
transforming as singlet under the Standard-Model gauge group); in
the other class they are felt also by gauge fields. In the former case,
the Standard-Model fields are confined to a (3+1)-dimensional subspace of a
higher-dimensional space some dimensions of which are compactified with a
relatively large radius. The absence of any observed deviation from ordinary
Newtonian gravity in Cavendish-type laboratory experiments implies that the
largest compactification radius is smaller than 
around 0.2 mm \cite{Hoy01}.

The main goal in both classes of models is to provide a unified theory in
which the electroweak scale $M_W \sim 10^2  \; 
{\rm GeV} $ and the high energy scales (Planck \cite{Ark98}, 
string \cite{Wit96},
and GUT scales \cite{Die98}) can coexist. The same scenario has also been
successfully applied to neutrino as well as to axion phenomenology. Namely,
a higher-dimensional seesaw mechanism may provide light neutrino masses
without heavy mass scales \cite{Die99}. 
Similarly, axion invisibility can be
achieved in extra dimensions even with a low fundamental Peccei-Quinn (PQ)
scale \cite{Die00}. 

The CERN Axion Solar Telescope (CAST) is designed
to search for solar axions of a broad energy spectrum which peaks at about
4 keV, through their conversion into real photons
inside the transverse magnetic field \cite{Zio99,Ira02}.
This telescope may improve the current laboratory bounds on the axion-photon
coupling, $g_{a\gamma\gamma} \alt 6 \times 10^{-10} \;{\rm GeV^{-1}}$
for $m \alt 0.03$ eV and $g_{a\gamma\gamma} \alt 6.8 - 10.9 \times
10^{-10} \; {\rm GeV^{-1}}$ for $ m \sim 0.05 - 0.27 \;{\rm eV}$
\cite{Mori98}, by a factor of ten or even more. It also has the potential
to extend for the first time the axion searches beyond the 
limit $g_{a\gamma\gamma} \alt 10^{-10} \; {\rm GeV^{-1}}$ 
arising from astrophysical constraints on anomalous 
energy loss by stars \cite{Raf02}. Although the CAST telescope could in 
principle be sensitive to axion masses in the range of a few keV, the
coherence-loss constraints \cite{Bib89,Laz92} reduce the sensitivity down to 
around 1 eV. 

The first goal of
the present note is to interpret prospects of CAST  
in the light of the theory with large extra spatial dimensions. 
We focus on the case when the limit on the 
size of two large extra compact dimensions is set by direct tests of gravity
\cite{Hoy01}. Our second goal is to explore the potential of CAST for testing
the presence of large extra dimensions. 

\section{\label{2}QCD AXIONS AND CAST}

Axions are pseudoscalars arising in models which
resolve the strong {\it CP} problem in quantum chromodynamics (QCD) by the
PQ mechanism \cite{Pec77}. Owing to their potential abundance
in the early universe, they are also well-motivated candidates for the
dark matter of the universe.
In both classes of (conventional) invisible axion models  referred to
as KSVZ or hadronic axion models \cite{Kim79} 
(where axions do not couple to electrons at tree level)
and DFSZ or grand unified theory
(GUT) models \cite{Din81}, the axion-photon coupling strength
is given by the relation
\begin{equation}
   g_{a\gamma\gamma} = \frac{\alpha}{2 \pi f_{\rm PQ}}\,
                       (E/N - 1.93 \pm 0.08)\;.
   \label{eq1}
\end{equation}
Here $E/N$ is a model-dependent numerical
parameter for hadronic axions, while for DFSZ axions $E/N = 8/3$.
Furthermore, the mass of the (QCD) axion $m_{\rm PQ}$ is related to the
PQ symmetry breaking scale $f_{\rm PQ}$ by
\begin{equation}
   m_{\rm PQ} = 6\,{\rm eV}\,
                         \frac{ 10\,^6}{f_{\rm PQ}/{1\,\rm GeV}}\;.
   \label{eq2}
\end{equation}
In order to avoid ambiguities owing to the model-dependence of the parameter
$E/N$ for hadronic axions, it proved more convenient to make constraints
on the axion-photon coupling than on the PQ energy scale or on the axion mass.

In contrast, cosmological considerations and astrophysical arguments (i.e., 
axion emission due to nucleon-nucleon bremsstrahlung from the supernova
SN 1987A) bound the axion mass into two possible ranges \cite{Raf96}. 
The first window is $10^{-5} \; {\rm eV} \alt m_{\rm PQ} \alt
10^{-2} \; {\rm eV}$, in which case the axion could constitute the
cold dark matter of the universe. The second one, being around ten to 
twenty electronvolts, appears to be of interest for hot dark matter.
However, such astrophysical constraints on $m_{\rm PQ}$, 
although the most stringent, suffer from statistical weakness (with only
19 neutrinos being observed) as well as from all uncertainties
related to the axion emission from a hot/dense medium. 
It is therefore of crucial importance to probe the axion properties
in a model-independent way \cite{Krc98,Krc01}. 

Currently, laboratory searches for solar axions \cite{Laz92,Avi98,Mori98,Ber01} 
are being extended by the
CAST experiment at CERN. This telescope uses a decommissioned LHC 
prototype magnet with a field
of 9 T and a length $L$ of 10 m. The magnet contains two straight beam pipes
with an effective cross sectional area $S = 2\times 14 \; {\rm cm^2}$, and
is mounted on a moving platform with
low-background X-ray detectors on either end
allowing it to track the Sun about 3 hours per day.

Hadronic axions could be produced abundantly in the core of the Sun by the
Primakoff conversion of the blackbody photons in the Coulomb fields of nuclei
and electrons in the solar plasma. The outgoing axion flux
is robust and does not depend on subtle details of the solar model.
It is approximately given by \cite{DiL00}
\begin{eqnarray}
   \frac{d\Phi(E)}{dE}&=&4.20 \times 10^{10} {\rm cm^{-2}\,s^{-1}\,keV^{-1}}
        \left(\frac{g_{a\gamma\gamma}}{10^{-10}\,{\rm GeV^{-1}}}\right)^2
    \nonumber \\
      && \times
      \frac{E\,p^2}{e^{E/1.1} - 0.7}\,(1 + 0.02m)\;.
      \label{eq3}
\end{eqnarray}
Here $d\Phi(E)/dE$ is the axion flux at the Earth,
differential with respect to axion energy ($E$),
and expressed as a function of axion mass ($m$).  
The quantities $E$, 
$p = \sqrt{E^2 - m^2}$, and $m$ are to be taken in keV. 
The probability for an axion-to-photon conversion  
in the presence of a transverse magnetic field ($B$) and a refractive
medium (i) is given by \cite{Bib89}
\begin{eqnarray}
 P\,^i_{a\rightarrow\gamma}(m)&=&\left(\frac{Bg_{a\gamma\gamma}}{2}\right)^2\,
                              \frac{1}{q_i^2 + \mu_i ^2/4}                     
    \nonumber \\
      && \times
    \left[1 + e^{-\mu_i L} - 2e^{-\mu_i L/2}{\rm cos}(q_iL)\right] ,
  \label{eq4}
\end{eqnarray}
where $q_i = |(m_{\gamma \,i}^2 - m^2)/2E|$ is the momentum difference between
photons in the medium  and axions, and $\mu_i$ denotes the inverse 
absorption length for X-rays. 
The effective mass (plasma frequency)
for an X-ray in He can be described in terms of the operating 
pressure $P_i$ (at 300 K) 
as $m_{\gamma \,i}/{\rm 1\,eV} \approx \sqrt{(P_i/{\rm 1\,atm})/15}\,$. 
The coherence condition $q_i\,L<\pi$ \cite{Laz92}
requires $m \alt 0.02~{\rm eV}$ for
a photon energy of 4.2 keV (the average axion energy) and a coherence length 
of 10 m in vacuum. To search for axions more massive, coherence can be restored
by filling the magnetic conversion region with buffer gas.   
Integrating over all axion energies, the expected
number of photons $N_{\gamma \,i}(m)$, being detected during the times of 
solar alignment with the magnet ($t_i$), is finally 
\begin{equation}
   N_{\gamma \,i}(m) = \int_{0}^{\infty}\frac{d\Phi(E)}{dE}\,
                P\,^i_{a\rightarrow\gamma }(m)\,S\,t_i\,dE\;,
   \label{eq5}
\end{equation}
assuming 100\% detection efficiency for the conversion X-rays.
At a fixed pressure $P_i$, the response of CAST will be a sharply peaked
function of the actual axion mass $m$, with the fractional resolution
$\Delta m/m_{\gamma\,i} \approx 5.2 \times 10^{-4} 
(m_{\gamma\,i}/1\,{\rm eV})^{-2}$.  
A general analysis of the experimental prospects \cite{Zio99}
explores the full two-dimensional ($m, g_{a\gamma\gamma}$) space for QCD 
axions rather than the narrow band defined by conventional axion models
(although it remains the best-motivated region), as it
is shown in Fig.~\ref{fig2}. The experiment is being operated in a scanning
mode in which the gas pressure is varied in appropriate steps 
(1\,yr with vacuum, an additional 1\,yr with a He gas pressure increased from
0-1\,atm in 100 increments, and an additional 1\,yr with 1-10\,atm in 365
increments) to cover a range of possible axion masses up to 0.82 eV. 

\section{\label{3}CAST AND LARGE EXTRA DIMENSIONS}

Large extra dimensions aim to stabilize the mass 
hierarchy (i.e., the
hierarchy between the Planck scale and the electroweak scale) by producing
the hugeness of the Planck mass $M_{\rm Pl}$ via the relation
\begin{equation}
      M_{\rm Pl}^{2} = M_{D}^{n+2} \, V_{n}\;,
      \label{eq6}
\end{equation}
where $V_n \equiv R^n$ is the full volume of the compactified space, and the
fundamental scale is set at $M_D \sim \; {\rm TeV}$. As already
stressed,
a singlet higher-dimensional axion field is also free to propagate into the
bulk and therefore a similar volume-suppressed formula can be used to lower
the fundamental Peccei-Quinn symmetry-breaking scale $\bar{f}_{\rm PQ}$
\cite{Ark98,Die00,Cha00}
\begin{equation}
      f_{\rm PQ}^2 = \bar{f}_{\rm PQ}^2 \, M_S^{\delta} \, V_{\delta}\;,
 \label{eq7}
\end{equation}
where $M_S $ is the string scale, $M_S \sim M_D $. Since the
phenomenologically allowed region for $f_{\rm PQ}$ (also generating the
coupling between the axion and matter) is such that $f_{\rm PQ} \ll 
M_{\rm Pl}$, the axion must be restricted to a subspace of the full
higher-dimensional bulk ($\delta < n$), if $\bar{f}_{\rm PQ}$ is to 
reside in the
TeV-range \cite{Die00}. Still, $\delta = n$ is possible for
$\bar{f}_{\rm PQ} \ll \; {\rm TeV}$ \cite{DiL00,Hor02}.

In Ref.~\cite{Die00} the full generalization of the higher-dimensional PQ
mechanism was given, including a thorough discussion of how extra space
dimensions may contribute to the invisibility of the PQ axion. All new
phenomena contributing to the invisibility of the axion and found there rely 
on a nontrivial axion mass matrix. Such a matrix is induced by a
mixing between the four-dimensional axion and the infinite tower of KK
excitations. The most interesting phenomenological consequence implied by
such a mixing  is a decoupling of the mass eigenstate of the axion from the
PQ scale (for $m_{\rm PQ} > (1/2)R^{-1}$). Since this feature is crucial 
for our
considerations here, we discuss it in more detail below.

\begin{figure}[ht]
\centerline{\includegraphics[width=80mm,angle=0]{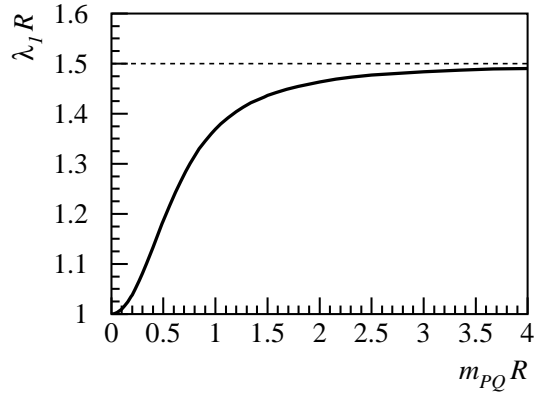}}
               \caption{The mass of the axion first-mode as a function of the
               dimensionless product $m_{\rm PQ}R$, where $m_{\rm PQ}$ is
               given by Eq.~(\ref{eq2}).}
      \label{fig1}
   \end{figure}

Now, we can focus on the higher-dimensional case by considering first the KK
decomposition of the axion field. As a major step, we need to calculate the
estimated number of X-rays at the pressure $P_i $ as a function of the KK
axion mass. The masses of the KK modes are given by
\begin{equation}
      m_{\vec{\delta}} = \frac{1}{R} \sqrt{ n_1^2 + n_2^2 + ...
                        + n_{\delta}^2}
                       \equiv \frac{|\vec{\delta}|}{R}\;,
    \label{eq8}
\end{equation}
where we assume that all $n$ extra dimensions are of the same size $R$.
When the mass splitting for the size $R$ ($\sim 1/R$) is sufficiently small,
one is allowed to use integration instead of summation \cite{Gui99}.
We have already mentioned that because of the nontrivial axion mass matrix,
neither the four-dimensional axion nor the KK states represent the mass
eigenstates. Instead, the eigenvalues are given as solutions to the
transcendental equation \cite{Die00}
\begin{equation}
    \pi R \lambda \,{\rm cot}(\pi R \lambda) = \frac{\lambda ^2}
                                             {m_{\rm PQ}^2}\;.
   \label{eq9}
\end{equation}
Hence, in order to estimate the number of modes with the KK index between $n$
and $n + dn$, one should parameterize the whole set of eigenvalues of 
(\ref{eq9}). This can be done by solving (\ref{eq9}) for two limiting 
cases, $m_{\rm PQ} R \gg 1$ and $m_{\rm PQ} R \ll 1$. 
We find for the eigenvalues 
\begin{eqnarray}
 && \lambda _0 \approx m_{\rm PQ}\;,  
    \nonumber \\
 && \lambda _n \approx \frac{n}{R}\;, \; \; n=1,2,...\;\;\; 
                                      {\rm if}\;\;\; m_{\rm PQ} R \ll 1\;,    
    \label{eq10}
\end{eqnarray}
and
\begin{equation}
 \lambda _n \approx \frac{2n+1}{2R}\;, \; \; n=0,1,2,...\;\;\;
                                         {\rm if}\;\;\; m_{\rm PQ} R \gg 1\;.
   \label{eq11}
\end{equation}
The results from Ref.\ \cite{Die00}, where only the mass of the axion 
zero-mode was estimated, can now be easily reproduced from our expressions
(\ref{eq10}) and (\ref{eq11}). In Fig.~\ref{fig1} we show the mass of the 
first KK state as a function of $m_{\rm PQ}R$. It can be seen how the
mass quickly approaches its limiting value (3/2)$R^{-1}$. A similar feature
was found in Ref.\ \cite{Die00} for the zero-mode.
Finally, the total number of X-rays due to all modes of the KK tower reads
\begin{eqnarray}
  &&N_{\gamma \,i}^{KK}=S_{\delta}R\,^{\delta} 
           \int_{0}^{\infty}dm\,m^{\delta -1} N_{\gamma \,i}(m)\,G(m) 
    \nonumber \\
            &&\;{\rm if}\;\; m_{\rm PQ} R \ll 1
    \label{eq12}
\end{eqnarray}
and 
\begin{eqnarray}
  &&N_{\gamma \,i}^{KK}=S_{\delta}R\,^{\delta} 
           \int_{0}^{\infty}dm\,m^{\delta -1} N_{\gamma \,i}(m+1/2R)\,
    \nonumber\\
           &&\times G(m+1/2R)
    \nonumber \\
            &&\;{\rm if}\;\; m_{\rm PQ} R \gg 1\;,
    \label{eq13}
\end{eqnarray}
where $S_{\delta} \equiv 2 \pi ^{\delta/2}/\Gamma (\delta /2)$ is the 
surface of a unit radius in $\delta$ dimensions and $G(m)$ is defined as
\begin{equation}
   G(m) = \tilde{m}^4\,(\tilde{m}^2+1+\pi^2/y^2)\,^{-2}\;,
\label{eq13a}
\end{equation}
with $\tilde{m} \equiv m/m_{\rm PQ}$ and $y \equiv 1/m_{\rm PQ}R$. The function
$G(m)$ arises from the mixing between the KK axion modes entering the KK
decomposition of the higher-dimensional axion field and the corresponding 
normalized mass eigenstates \cite{Die00}. It also implies both production
and detection of KK axions to occur on our Standard-Model brane. 
The function $G(m)$ also incorporates the effect of rapid 
decoherency \cite{Die00} 
of the only linear combination of KK states of the bulk axion which couples
to Standard-Model fields. This means that the production and subsequent
detection of this particular linear combination of KK states are strongly
suppressed. As a consequence our results always reflect a volume-suppressed
coupling $g_{a\gamma\gamma} \sim 1/f_{\rm PQ}$. If it was not for the
decoherency, the linear combination would be coupled to photons with an
unsuppressed coupling $1/{\bar f}_{\rm PQ}$.

\section{\label{4}DISCUSSION}

\begin{figure}
   \centerline{\includegraphics[width=80mm,angle=0]{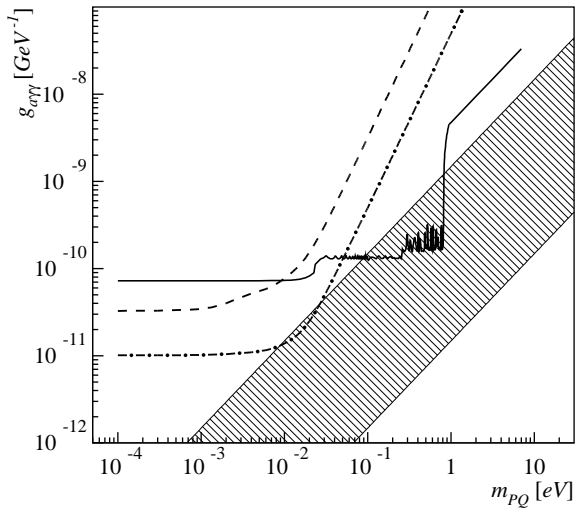}}
      \caption{
               Limits on the axion-photon coupling $g_{a\gamma\gamma}$ as
               a function of the fundamental PQ mass $m_{\rm PQ}$. 
               The solid line, corresponding
               to prospects of CAST for QCD axions, is obtained 
               using numerical values from Ref.~\cite{Zio99}.
               The dashed region marks the theoretically favored relation
               between $g_{a\gamma\gamma}$ and $m_{\rm PQ}$ in axion models
               in four dimensions. 
               The dashed and dot-dashed
               line correspond to prospects of CAST for KK axions in
               the case of two extra dimensions ($R=0.150$ mm) for
               $\delta = 1$ and for $\delta = 2$, respectively.}
   \label{fig2}
   \end{figure}

In order to achieve an upper limit on the coupling of the axion to photons  
from the prospects of CAST in the framework of large extra dimensions,  
we apply the central limit theorem at 3\,$\sigma$ level 
\begin{equation}
    \sum_i N_{\gamma \,i}^{KK} \alt 3\; \sqrt{\sum_i N_{b\,i}} \;,
    \label{eq14}
\end{equation}
where $N_{b\,i}$ is the background of the X-ray detector (with numerical
values taken from Ref.~\cite{Zio99}). For the sake of simplicity, it is  
assumed that all axions have an average energy of 4.2 keV. 
Combining Eq.~(\ref{eq14}) with
Eqs.~(\ref{eq3})-(\ref{eq5}) and (\ref{eq12})-(\ref{eq13a}) for the case
of two extra dimensions (we take the largest compactification radius  
of 0.150 mm as set by direct tests of Newton's law \cite{Hoy01}), 
we have derived limits on the axion-photon coupling $g_{a\gamma\gamma}$ as a
function of the fundamental PQ mass, as shown in Fig.~\ref{fig2}.
Although the multiplicity of KK
states to which CAST could be sensitive is large ($\sim 10^3$ for $\delta = 1$
and $\sim 10^6$ for $\delta = 2$), 
one can see from Fig.~\ref{fig2} that the upper limit
on $g_{a\gamma\gamma}$ is only at most
an order of magnitude more stringent 
than that obtained in conventional theories. 
This is due to the fact that CAST is a tuning experiment, i.e., the coherence 
condition at a fixed pressure is fulfilled only within a very narrow window 
of axion masses around $m\approx m_{\gamma\,i}$. The corresponding width  
ranges (depending on pressure) from $\sim 10^{-2}\,{\rm eV}$ down to 
$\sim 10^{-3}\,{\rm eV}$. Another feature visible in Fig.~\ref{fig2}
is a strong decrease in sensitivity to $g_{a\gamma\gamma}$ for 
$m_{\rm PQ} R \agt 20$ if $\delta = 2$ and for somewhat lower values if
$\delta = 1$. This is due to the fact that in the regime $m_{\rm PQ} R \gg 1$,
$G(m)$ decreases as fast as $1/(m_{\rm PQ} R)^8$. It is just the regime in 
which the obtained limits on $g_{a\gamma\gamma}$ cannot be coupled with 
the zero-mode axion mass ($m_a\approx(1/2)R^{-1}=6.6\times 10^{-4}\,{\rm eV}$)
via relations (\ref{eq1})
and (\ref{eq2}) because in higher dimensions the mass of the axion is
approximatively given as $m_a \approx {\rm min}
((1/2)R^{-1}, m_{\rm PQ})$ \cite{Die00}.
In contrast with the case of ordinary QCD axions, in theories with large
extra dimensions zero-mode axions with masses outside the
favored band (as determined by conventional axion models in four dimensions)
arise quite naturally.

Now we would like to point to a new phenomenon predicted for CAST: 
sensitivity to particular KK axions. We have already noted that physical 
KK modes are given by Eqs.~(\ref{eq10}) and (\ref{eq11}).  
It is expected that more than one axion signal  
may be observed at different pressures of the gas. Therefore, the detection
of the corresponding X-rays at least at two pressures may be the signal for
the presence of large extra dimensions. As the CAST experiment is scanning 
the range of axion masses up to 0.82 eV, this requirement
actually defines a sensitivity of the experiment to test the compactification
radius. From Eqs.~(\ref{eq10}) and (\ref{eq11}) we obtain 
$R\agt 370\,{\rm nm}$ if $m_{\rm PQ} R > 1$ and
$R\agt 250\,{\rm nm}$ if $m_{\rm PQ} R < 1$, with 
$g_{a\gamma\gamma}\sim 10^{-10}\,{\rm GeV}^{-1}$.
It should be noted here that with the CAST sensitivity to $g_{a\gamma\gamma}$
the former result holds only for $m_{\rm PQ} R \agt 1$; for
$m_{\rm PQ} R \gg 1$ the sensitivity rapidly decreases due to the suppression
from the $G(m)$ function.   
The present modifications of CAST may increase its sensitivity
to $g_{a\gamma\gamma}$ by a factor of 1.5 \cite{Ira02}, providing 
the sensitivity 
as mentioned above is of the order of that derived from the solar age 
consideration \cite{Krc03}. Note that in a recent review
of the Particle Data Group \cite{Hag02} the 
bound on $R$ for the case
of two extra dimensions, coming from astrophysics, was listed to have a value 
within the range 90 to 660 nm (for the most stringent
constraints, see the recent work \cite{Han03}).

In conclusion, we have explored the potential of the CAST experiment for
observing KK axions coming from the solar interior. Because of the
restrictive coherence condition, in theories with  two extra dimensions (with
$R=0.150$~mm) a sensitivity in axion-photon coupling 
improves at most one order of magnitude in both data taking phases. 
In this case, the
obtained limit on $g_{a\gamma\gamma}$ cannot be coupled
with the mass of the axion,
which is essentially given by the (common) radius of the extra dimensions.
In addition, we have demonstrated that the CAST experiment, being a tuning
experiment with respect to axion masses, may not be sensitive only to an
integrated effect of KK modes up to the kinematical limit but also to
particular KK axions. With a requirement to have at least two signals while
changing pressure of the gas, we have found that CAST is capable of 
probing (two) large extra dimensions with a
compactification radius $R$ down to around 250 nm if
$m_{\rm PQ} < 1/(2R)$,
and down to around 370 nm if $1/(2R) < m_{\rm PQ}$.


\begin{thebibliography}{100}
\bibitem{Krc03}
   By simply scaling the solar age limit of QCD axions
   ($g_{a\gamma\gamma}\alt 2.4\times10^{-9}\,{\rm GeV}^{-1}$) \cite{Raf96}
   with $(R\;m_{\rm max})^{-\delta/2}$ due to
   the multiplicity of KK modes and observing that an allowed maximum mass is
   $m_{\rm max} \sim {\rm keV}$ before the solar flux gets suppressed by 
   the kinematic threshold, one obtains for two extra dimensions:  
   (i) for $R = 0.150$ mm:
   $g_{a\gamma\gamma}\alt 3\times 10^{-12}\,{\rm GeV}^{-1}$ if $\delta = 1$ 
   and
   $g_{a\gamma\gamma}\alt 3\times 10^{-15}\,{\rm GeV}^{-1}$ if $\delta = 2$, 
   and
   (ii) for $R = 250$ nm:
   $g_{a\gamma\gamma}\alt 7\times 10^{-11}\,{\rm GeV}^{-1}$ if $\delta = 1$ 
   and
   $g_{a\gamma\gamma}\alt 2\times 10^{-12}\,{\rm GeV}^{-1}$ if $\delta = 2$.
\end{thebibliography}
\end{document}